# Comment on "Comment on 'On Visibility in the Afshar Two-Slit Experiment' "

R. E. Kastner
Feb. 21, 2010

Abstract. Tabish Qureshi (2010) has recently objected to an aspect of my discussion of a thought experiment by Srikanth (2001). I believe his objection is based on a misunderstanding about my presentation, but I accept responsibility for not being more clear. Here I attempt to clear up any confusion and to point out that one can indeed obtain interference in this experiment based on taking care to correctly characterize the detection process. I also emphasize that Qureshi and I are in full agreement concerning the assertion that when interference is present, there is no genuine "which slit" information, and that in this case, the measurement necessary to preserve interference precludes determination of an actual slit of origin for the photon.

Srikanth (2001) proposed an "enhanced detection screen" for a two-slit experiment, in which each element of the detector, besides indicating position x on the screen, contains a hypothetical "vibrational" component with states $\{|v_a\rangle, |v_b\rangle\}$ which could couple to the particle's momentum based on slit of origin (slit *a* or *b*). Unfortunately, Dr. Qureshi does not seem to have noticed that we are in full agreement concerning the lack of which-way information in this experiment when interference is present. It is true that one must take special care to preserve interference in the Srikanth experiment since, as Qureshi notes, merely calculating the probability corresponding to the usual observable for the firing of a detector at a location x on the screen ($|\phi_x\rangle\langle\phi_x|$) will not give interference, given the orthogonality of the vibrational states. The enhanced detection setup suggested by Srikanth, in order to preserve interference, must constitute a measurement of the *total* detector observable ($|\phi_x\rangle\langle\phi_x| \otimes |\phi_v\rangle\langle\phi_v|$, where

$|\phi_v> = \frac{1}{2} |v_a> + |v_b>$ (a superposition of the "vibrational component" states), in order to preserve interference. Then, of course, the vibrational component of the detector will not contain any which-way information, which is what I argued in my (2009). Thus Qureshi and I are in full agreement that, when interference is present, one cannot claim to have 'which-slit' information, even with an enhanced detector that superficially appears to be able to provide that information. My discussion of Srikanth's experiment was in the context of the Afshar experiment (e.g., 2007) which also seems to provide a post-selection "which-slit" measurement but which, similarly, does not really provide which-slit information for analogous reasons (i.e., the post-selection just 'collapses' a pre-existing superposition of slits analogous to the state $|\phi_v>$). Srikanth also argued that one would not get genuine 'which-slit' information from his experiment, remarking that such a detection process would "leave behind a remnant superposition" of the vibrational states. This is reflected in the measured vibrational state $|\phi_v>$.